\documentclass[aps,prc,twocolumn,superscriptaddress]{revtex4}

\usepackage{epsfig}
\begin{document}

\title{Experimental investigation on the temperature dependence of the nuclear level density parameter}

\author{Balaram Dey}
\affiliation{Variable Energy Cyclotron Centre, 1/AF-Bidhannagar, Kolkata-700064, India}

\author{Deepak Pandit}
\affiliation{Variable Energy Cyclotron Centre, 1/AF-Bidhannagar, Kolkata-700064, India}

\author{Srijit Bhattacharya}
\affiliation{Department of Physics, Barasat Govt. College, Barasat, N 24 Pgs, Kolkata - 700124, India}

\author{K. Banerjee}
\affiliation{Variable Energy Cyclotron Centre, 1/AF-Bidhannagar, Kolkata-700064, India}

\author{N. Quang Hung}
\affiliation{School of Engineering, Tan Tao University, Tan Tao University Avenue, Tan Duc Ecity, Duc Hoa, Long An Province, Vietnam}

\author{N. Dinh Dang}
\affiliation{Theoretical Nuclear Physics Laboratory, RIKEN Nishina Center for Accelerator-Based Science,
RIKEN, 2-1, Hirosawa, Wakocity, Saitama 351-0198,Japan and Institute for Nuclear Science and Technique, Hanoi, Vietnam}

\author{Debasish Mondal}
\affiliation{Variable Energy Cyclotron Centre, 1/AF-Bidhannagar, Kolkata-700064, India}

\author{S. Mukhopadhyay}
\affiliation{Variable Energy Cyclotron Centre, 1/AF-Bidhannagar, Kolkata-700064, India}

\author{Surajit Pal}
\affiliation{Variable Energy Cyclotron Centre, 1/AF-Bidhannagar, Kolkata-700064, India}

\author{A. De}
\affiliation{Department of Physics, Raniganj Girls' College, Raniganj-713358, India}

\author{S. R. Banerjee}
\email[e-mail:]{srb@vecc.gov.in}
\affiliation{Variable Energy Cyclotron Centre, 1/AF-Bidhannagar, Kolkata-700064, India}


\date{\today}

\begin{abstract}
The effect of temperature (T) and angular momentum (J) on the inverse level density parameter (k) has been studied by populating the compound nucleus $^{97}$Tc in the reaction $^{4}$He + $^{93}$Nb at four incident beam energies of 28, 35, 42 and 50 MeV. For all the four energies, the value of k decreases with increasing J. The T dependence of k has been compared for two angular momentum windows with different theoretical predictions as well as with FTBCS1 calculation which takes into account the quasiparticle-number fluctuations in the pairing field. Interestingly, the experimental data are in good agreement with the theoretical calculations at higher J but deviate from all the calculations at lower J.   

\end{abstract}
\pacs{25.70.Jj, 25.70.Gh, 24.10.Pa}
\maketitle

\section{Introduction}
The nucleus is a many body quantum system which experiences many different configurations even when fairly small excitation energy is provided. It is now a very well-known fact that the density of quantum mechanical states increases rapidly with excitation energy and soon becomes very large. As a result, the nucleus leaves the discrete region and enters the region of quasi-continuum and continuum. Due to this complexity, statistical concept and models are not only appropriate but also crucial for comprehension and prediction of various nuclear phenomena. Intriguingly, the nuclear level densities are indispensable in the study of nuclear reaction cross sections, nuclear reaction rates \cite{roch} which are needed for astrophysical calculations (inputs in modeling stellar evolution and nucleosynthesis), fission or fusion reactor design, transmutation of nuclear waste and production of radioactive isotopes in therapeutic uses in nuclear medicine. Along with that, they also provide information about the thermodynamic quantities such as temperature and entropy as well as pairing correlations in the nuclear structure study \cite{melby}. The knowledge of nuclear level density is also very crucial while extracting the parameters of the giant dipole resonance built on highly excited states of the nuclei\cite{balaram,srijit}. However, the characterization of the nuclear level density (NLD) in the regions of high excitation energy, angular momentum and different nuclear shapes are in large part phenomenological. The most commonly used analytical expression for calculating NLD is based on the work of Bethe \cite{bethe} for a system of non-interacting fermions. The backshifted Fermi gas model of NLD for a spherical nucleus of mass number A at excitation energy E$^*$ and angular momentum J is given by

\begin{equation}
\rho (E^*,J) = \frac{2J+1}{12I^{3/2}} \sqrt{a} \frac{\exp{(\sqrt{(aU)}})}{U^2}. \label{eqn1}
\end{equation}

Here U = E* - E$_{rot}$ - $\delta$p is the available thermal energy. E$_{rot}$, $\delta$p and I are the energy bound in the rotation, paring energy and effective moment of inertia, respectively. The nuclear level density parameter ($a$) is related to the single particle level density in the region of the Fermi energy and is correlated to the mass of the nucleus as $a$ = A/k, where k is the inverse level density parameter. The temperature dependence of the NLD parameter has been investigated by various theoretical approaches \cite{shlomo, lestone, mugh, prakash, jnde} and experimental methods \cite{hagel,gonin,chibi,fine,fabris}. It is found from the experimental analysis of nuclear resonances and evaporation spectra that for cold nuclei k $\sim$ 8 MeV but for high temperatures like T = 5 MeV, k is around 13 MeV. The understanding of this behavior has been attempted by taking into account the effects of correlation, i.e. the T dependence of the frequency dependent effective mass as well as the finite size effect, the momentum dependence of the effective mass, the effects of continuum and the shell effects \cite{shlomo,prakash,jnde}.

It may be noted that the level density formalism given in Eq.(1) is for a system of non-interacting fermions with equidistant single particle states and does not include the collective enhancement of NLD due to the coupling of rotational as well as vibrational degrees of freedom with the single-particle degrees of freedom. The enhanced level density is expressed as $\rho$(E*, J) = $\rho_{int}$(E*, J) K$_{coll}$(E*), where K$_{coll}$(E*) is the collective enhancement factor consisting of both vibrational and rotational contributions \cite{coll_enhance}. Very recently, neutron evaporation spectra from $^{201}$Tl*, $^{185}$Re*, and $^{169}$Tm* compound nuclei have been measured at two excitation energies (E* $\sim$ 37 and 26 MeV) to study the effect of collectivity by extracting the inverse level density parameter (k) \cite{proy1}. It was observed that for large ground state deformed nuclei ($^{185}$Re* and $^{169}$Tm*) the value of k decreased substantially at the lower excitation energy, while for near spherical nucleus ($^{201}$Tl) it remains the same at the two excitation energies. The results indicated towards the strong correlation between collectivity and ground-state deformation. 

Recently, there have been ample of experimental efforts in order to comprehend the spin dependence of the level density parameter. In few measurements of angular momentum gated neutron evaporation spectra in A $\sim$ 119, 97 and 62, it was seen that the k value decreased with increase in J which indicated that level density increases with J \cite{kban1,proy2}. On the other hand, the inverse level density parameter extracted from the alpha evaporation spectra in A $\sim$ 180 and A $\sim$ 120 mass region showed that the value of k is either constant or increases with increase in angular momentum \cite{ykgupta1, ykgupta2}. However, theoretical calculation for similar masses shows that the k value should increase with increase in angular momentum for all the systems \cite{m_agar}. Interestingly, the $\gamma$-multiplicity gated proton spectra in A $\sim$ 105 showed a drastic dependence on fold. The spectra acquired a broad structure at higher multiplicity folds for proton energies beyond $\sim$15 MeV which was explained by a prescription of a localised enhancement of NLD \cite{mitra}. Thus, extremely exciting but conflicting experimental results on the spin dependence of the level density parameter motivate one to carry out further investigations.

In this work we report on the angular momentum gated neutron evaporation spectra at different excitation energies (30-50 MeV) for the reaction $^{4}$He + $^{93}$Nb. The specific advantage of using light ion reaction is that the major residues are of similar nature and in our case they are $^{95}$Tc, $^{94}$Tc and $^{93}$Tc depending on the excitation energy of the compound nucleus. However, for excitation energies above 42 MeV, another channel contributes ($\sim$20$\%$) due to ($\alpha$,2n) populating $^{91}$Nb. Nevertheless, the deformations of all the nuclei populated in the decay chain are similar and are of the order of beta $\sim$ 0.05.  The shell effects are also very small and similar. 
Therefore, the neutron evaporation spectra will have the contribution from similar kind of nuclei only.

\begin{figure}
\begin{center}
\includegraphics[height=8.0 cm, width=8.0 cm]{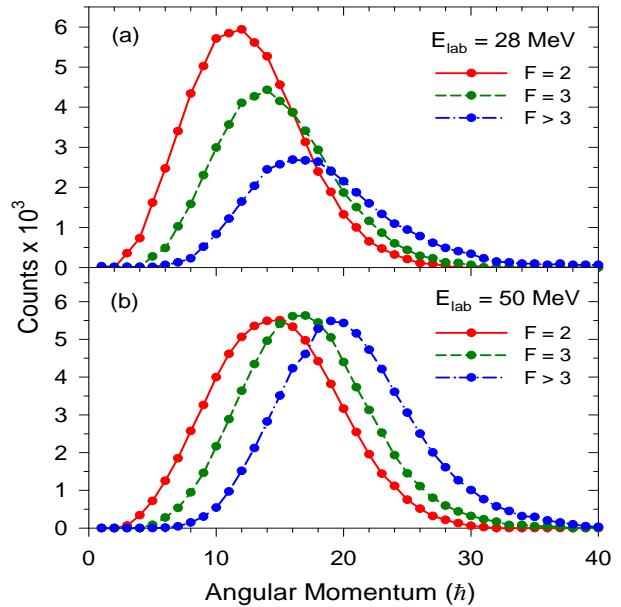}
\caption{\label{ang_Tc} Angular momentum distribution for different folds for the $^{4}$He+$^{93}$Nb system at 28 and 50 MeV incident energy.}
\end{center}
\end{figure}

\begin{figure}
\begin{center}
\includegraphics[height=12.0 cm, width=8.3 cm]{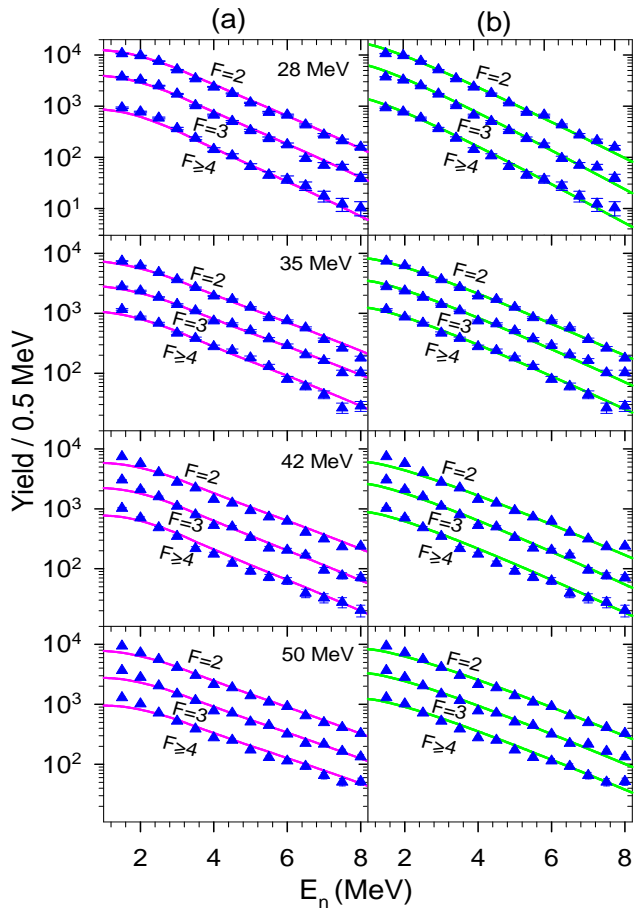}
\caption{\label{neut_Tc} (a) Neutron evaporation energy spectra (filled triangles) along with the CASCADE predictions (continuous line) for different folds (F) at incident energies of 28, 35, 42 and 50 MeV for $^{4}$He+$^{93}$Nb system. (b) Neutron evaporation energy spectra (filled triangles) along with the Maxwellian fitting (continuous line) for different folds (F) at incident energies of 28, 35, 42 and 50 MeV for $^{4}$He+$^{93}$Nb system. F=2 and F=3 data have been multiplied by 100 and 10, respectively.}
\end{center}
\end{figure}

\section{Experimental Details}
The experiment was performed at the Variable Energy Cyclotron Centre, Kolkata using alpha beam from the K-130 cyclotron.  A self supporting 1 mg/cm$^{2}$ thick target of 99.9$\%$ pure $^{93}$Nb target was used. Four different beam ($^4$He) energies of 28, 35, 42 and 50 MeV were used to populate the compound nucleus $^{97}$Tc at the excitation energies of 29.3, 36.0, 43.0 and 50.4 MeV, respectively. The maximum populated angular momenta for fusion were 16, 18, 19 and 20 $\hbar$, respectively. The evaporated neutrons from the compound nucleus were detected by a liquid organic scintillator (BC501A) based neutron detector \cite{kban} that was placed at a distance 1.5 m from the target position and at an angle of 90$^0$ to the beam axis.
Along with the BC501A neutron detector, a 50 element low energy $\gamma$-multiplicity filter \cite{deep} was also used to estimate the angular momentum populated in the compound nucleus as well as to get a fast start trigger for neutron time-of-flight (TOF) measurement. The multiplicity filter was split into two blocks of 25 detectors each, in a staggered castle type geometry to equalize the solid angle for each multiplicity detector element, and placed at a distance of 5 cm above and below the centre of the target. The efficiency of the multiplicity set-up was 56$\%$ as calculated using GEANT4 simulation.
A level-1 trigger (A) was generated from the multiplicity filter array when at least one detector each from the top and bottom blocks fired in coincidence above a threshold of 250 keV. Another trigger (B) was generated when the signal in BC501A detector crossed a threshold of 250 keV. An on-line coincidence of these two triggers (A and B) ensured the selection of neutron events and rejected the backgrounds. The TOF technique was employed for neutron energy measurement.  The neutron-$\gamma$ ray discrimination was achieved by both pulse shape discrimination (PSD) and TOF techniques. 
To keep the background of the detectors at a minimum level, the beam dump was heavily shielded with lead bricks and borated paraffin in both the experiments. A CAMAC electronics and VME based data acquisition system were used to simultaneously record the energy and time information of the detectors.

\begin{figure}
\begin{center}
\includegraphics[height=7.5 cm, width=8.3 cm]{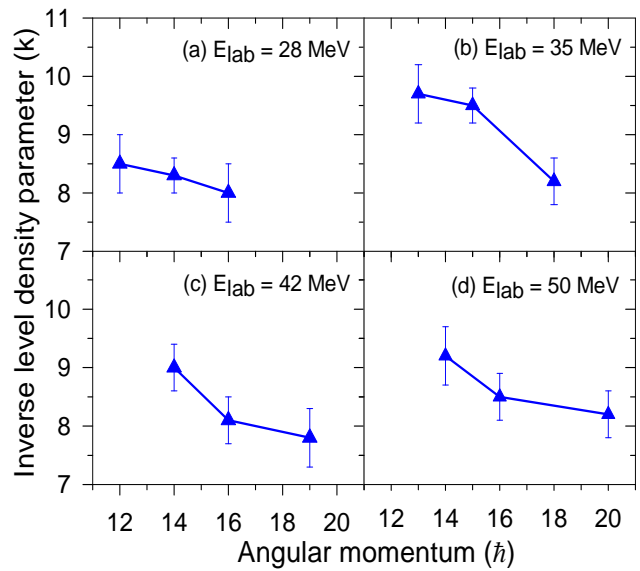}
\caption{\label{neut_ang} Angular momentum dependence of nuclear level density parameter at four incident energies.}
\end{center}
\end{figure}

\section{Data Analysis and Results}

The TOF spectrum was converted to energy spectrum using the prompt $\gamma$ peak as a time reference. The efficiency corrrection for BC501A neutron detector was performed using GEANT4 simulation \cite{geant}. In order to extract the inverse level density parameter from the experimental neutron spectra, the theoretical neutron energy spectra were calculated employing the statistical model code CASCADE \cite{cas}. The level density parameter prescription of Ignatyuk et al. \cite{ign1} was adopted which takes into account the nuclear shell effects at low excitation energy and connects smoothly to the liquid drop value at high excitation energy. The transmission coefficients for statistical calculation were obtained from the optical model. The potential parameters for neutron, proton and alpha were taken from \cite{per1}, \cite{per2} and \cite{sat}, respectively. The experimental fold distribution measured using the 50-element $\gamma$-multiplicity filter was converted to the spin distribution using GEANT4 simulation applying the approach discussed in Ref \cite{deep}. The simulated spin distributions deduced from the experimental fold distributions were used as inputs for different folds. The angular momentum distributions for different folds at 28 and 50 MeV incident energies are shown in Fig \ref{ang_Tc}. The moment of inertia of the CN was taken as I$_{eff}$ = I$_0$*(1 + $\delta_1$J$^2$ + $\delta_2$J$^4$) where I$_0$ is the moment of inertia of spherical nucleus. The role of the deformation parameters $\delta_1$ and $\delta_2$ were found to be inconsequential and the shape of neutron energy spectra was mostly dependent on the inverse level density parameter. The value of k has been extracted from the best fit statistical model calculations using a $\chi^2$ minimization in the energy range of 3$-$7 MeV (Fig \ref{neut_Tc}). The extracted inverse level density parameters are given in Table $-$ I for different angular momenta and excitation energies. 

\begin{figure}
\begin{center}
\includegraphics[height=8.0 cm, width=8.0 cm]{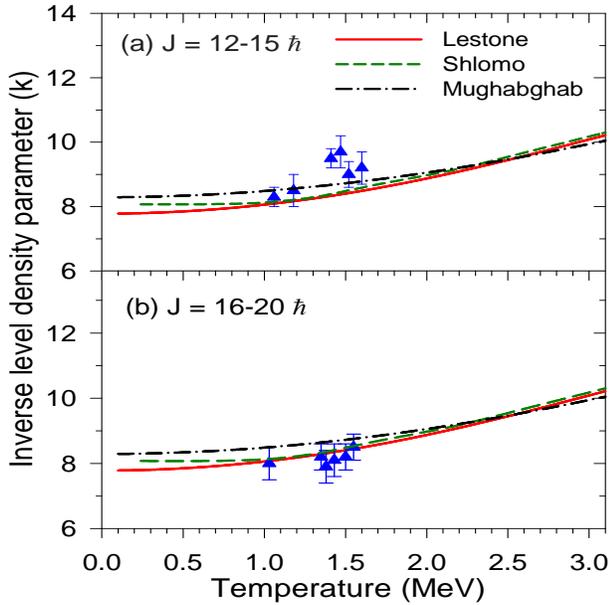}
\caption{\label{shlomo} Temperature dependence of nuclear level density parameter compared with different theoretical calculation. Green dashed line represents the calculation by  Shlomo, red dashed line represents the calculation by Lestone and black dot-dashed line represents the calculation by Mughabghab.}
\end{center}
\end{figure}

\begin{figure}
\begin{center}
\includegraphics[height=8.0 cm, width=8.0 cm]{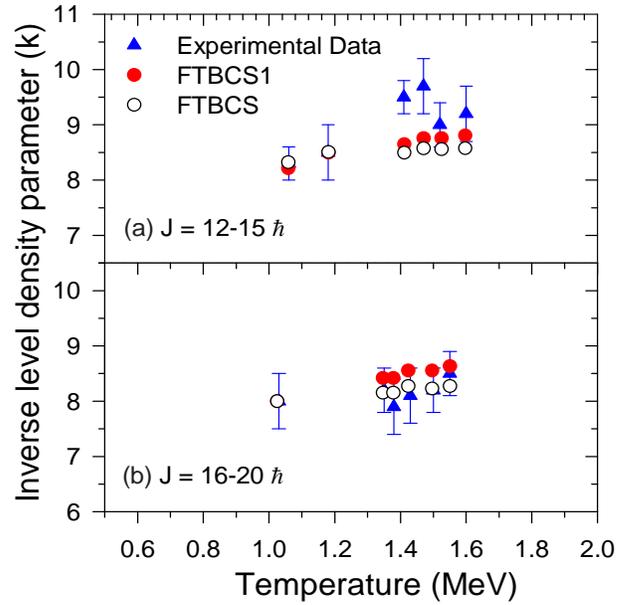}
\caption{\label{bcs} Temperature dependence of the nuclear level density parameter compared with the calculations of FTBCS and FTBCS1. }
\end{center}
\end{figure}


\begin{table}
\caption{\label{data} Average temperature and average angular momentum along with level density parameter for $^{4}$He+$^{93}$Nb system at different beam energies.}
\vspace{5mm}
  
\begin{center}		
\begin{tabular}{|c|c|c|c|c|}
\hline

Systems            &E$^{*}$  &$\;$$\left\langle J \right\rangle$$\;$& $\left\langle T \right\rangle$ &  $\;$ A/k $\;$  \\
                   &MeV      &$\hbar$&       MeV      &MeV$^{-1}$ \\ \hline
                   &         & 12$\pm$ 5    &$\;$ 1.18   $\;$ &  A/(8.5$\pm$ 0.5)  \\
$^{4}$He+$^{93}$Nb &29.3     & 14$\pm$ 6    &$\;$ 1.06   $\;$ &  A/(8.3$\pm$ 0.3)  \\     
@E$_{lab}$ = 28 MeV&         & 16$\pm$ 5    &$\;$ 1.03   $\;$ &  A/(8.0$\pm$ 0.5)  \\ \hline
                   &         & 13$\pm$ 4    &$\;$ 1.47   $\;$ &  A/(9.7$\pm$ 0.5)  \\
$^{4}$He+$^{93}$Nb &36.0     & 15$\pm$ 5    &$\;$ 1.41   $\;$ &  A/(9.5$\pm$ 0.3)  \\
@E$_{lab}$ = 35 MeV&         & 18$\pm$ 5    &$\;$ 1.35   $\;$ &  A/(8.2$\pm$ 0.4)  \\ \hline
                   &         & 14$\pm$ 5    &$\;$ 1.52   $\;$ &  A/(9.0$\pm$ 0.4)  \\
$^{4}$He+$^{93}$Nb &43.0     & 16$\pm$ 5    &$\;$ 1.43   $\;$ &  A/(8.1$\pm$ 0.4)  \\
@E$_{lab}$ = 42 MeV&         & 19$\pm$ 6    &$\;$ 1.38   $\;$ &  A/(7.8$\pm$ 0.5)  \\ \hline
                   &         & 14$\pm$ 5    &$\;$ 1.60   $\;$ &  A/(9.2$\pm$ 0.5)  \\
$^{4}$He+$^{93}$Nb &50.4     & 16$\pm$ 5    &$\;$ 1.55   $\;$ &  A/(8.5$\pm$ 0.4)  \\
@E$_{lab}$ = 50 MeV&         & 20$\pm$ 5    &$\;$ 1.50   $\;$ &  A/(8.2$\pm$ 0.4)  \\ \hline

\hline
\end{tabular}
\end{center}		
\end{table}

It is very interesting to note that the values of inverse level density parameter decrease with increase in angular momentum for all the incident energies (Fig \ref{neut_ang}). The result is in contrast to the theoretical calculations obtained under the framework of statistical theory of hot rotating nuclei which predicts that k should increase with increasing angular momentum \cite{m_agar}. However, the trend of the angular momentum dependence measured in this work is consistent with the previous measurements for $^{62}$Zn \cite{proy2} and $^{119}$Sb \cite{kban1}. The inverse level density parameter as a function of temperature, for two angular momentum windows, has also been compared with
the different theoretical calculations predicted by Shlomo \cite{shlomo}, Lestone \cite{lestone} and Mughaghab \cite{mugh} (Fig \ref{shlomo}).
It should be mentioned that the prediction of Shlomo has been taken for A $=$ 110 mass region \cite{shlomo} whereas the other two predictions have been calculated for A$=$97 mass using the formula given in Ref \cite{lestone, mugh}. The average temperatures of the populated compound nucleus for different excitation energies were calculated by fitting the experimental data with Maxwellian function ($\sqrt{E}exp(-E/T)$). 
It needs to be mentioned that none of the theoretical predictions include J effect. As can be seen from Fig \ref{shlomo}, the inverse level density parameter increases with increase in temperature for all the predictions. However, the data at higher angular momenta match very well with the prediction but are at disparity at lower angular momenta. This outcome was unexpected since the theoretical predictions do not include J effect and, thus, experimental data and theoretical calculation should have matched at lower J. The results indicate that the level density is suppressed at lower angular momenta. This deviation cannot be attributed to the effect of collectivity because the level density is not enhanced but suppressed compared to the Fermi gas model.

Recently, the temperature dependence of the level density was studied in hot medium-mass nuclei, which undergo a non-collective rotation about the symmetry axis \cite{dang}. The numerical calculations within the finite temperature BCS (FTBCS) and FTBCS1 theories have shown the pairing re-entrance in the pairing gap at finite angular momentum M (M is the z-projection of total angular momentum) and temperature T. The FTBCS1 theory includes the effects due to quasiparticle-number fluctuations in the pairing field and the z projection of angular momentum at T $\neq$ 0 MeV. The pairing re-entrance changes the T dependence of the level density from a convex function to a concave one. Similar calculation was performed to observe wheather the pairing re-entrance plays any important role in the case of $^{97}$Tc. It was observed that the FTBCS gaps collapse at a certain critical temperature, whereas the FTBCS1 gaps do not. 
Starting from M$=$12 $\hbar$ the FTBCS1 proton gap shows the pairing re-entrance effect, that is the pairing gap reappears and remains finite at T$>$0.3 MeV and M $\geq$ 12 $\hbar$. The value of k was extracted from the excitation energies using the relations k = 4AE$^*$/S$^2$. Since, the nucleus $^{97}$Tc is not spherical but slightly prolate and the calculations were carried out under the assumption of a spherical nucleus for which the z-projection M of the total angular momentum J coincides with J, the results of theoretical calculations have been re-normalized to match the corresponding data point at the lowest J for both the selected angular momentum window. As can be seen from Fig \ref{bcs}, the increase of k with T is observed in the data are also reproduced by the results of theoretical calculations although the latter agree better with the data for the higher J window whereas for the lower J window the theory underestimates the two data points at T = 1.41 MeV (J=15$\hbar$) and 1.47 MeV (J=13$\hbar$). The FTBCS and FTBCS1 predict similar results indicating that pairing re-entrance does not have much effect on the inverse level density parameter in this case. As could be discerned, collectivity and pairing re-entrance are not responsible for decrease in k with increase in angular momentum. Moreover, it may be noted that the effect of angular momentum on the k is not observed for higher masses \cite{ykgupta1, ykgupta2, man} but is only apparent in low and medium mass A $\leq$ 120 nuclei \cite{kban1, proy2}. More experimental data at both high and low mass region are required to understand this behavior.

\section{Summary and Conclusions}

Angular momentum gated neutron evaporation spectra have been measured in the reactions 
$^{4}$He + $^{93}$Nb at E$_{lab}$ $=$ 28, 35, 42 and 50 MeV to study the T and J dependence of the inverse level density parameter. It was observed that the inverse level density parameter decreased with increasing J for all the excitation energies. The T dependence of the inverse level density parameter was studied by selecting two angular momentum windows and compared with different theoretical calculations. Intriguingly, the results of all the theoretical calculations agree well with the data at higher J window but differ from the experimental data at lower J window. 

\section*{Acknowledgements}
The authors greatfully acknowledge helpful discussion with Professor J. N. De. The theoretical calculations within FTBCS and FTBCS1 were carried out using RIKEN Integrated Cluster of Clusters (RICC). N Quang Hung acknowledges the support by the National Foundation for Science and Technology Development (NAFOSTED) of Vietnam through Grant No. 103.04-2013.08.


\begin{thebibliography}{99}


\bibitem{roch}   T. Rauscher, Friedrich-Karl Thielemann,Karl-Ludwig Kratz, Phys. Rev. C 56, 1613 (1997).
\bibitem{melby}  E. Melby et., Phys. Rev. Lett 83, 3251 (1999).
\bibitem{balaram} Balaram Dey et al. , Phys. Lett. B 731, 92 (2014).
\bibitem{srijit} Srijit Bhattacharya et al., Phys. Rev. C 90, 054319 (2014).
\bibitem{bethe}  H. A. Bethe, Phys. Rev. 50, 332 (1936); Rev. Mod. Phys. 9, 69 (1937).
\bibitem{shlomo} S. Shlomo and J. B. Natowitz, Phys. Lett. B 252, 1987 (1990), Phys. Rev. C 44, 2878 (1991).
\bibitem{lestone} J. P. Lestone, Phys. Rev. C 52, 1118 (1995).
\bibitem{mugh} S.F.Mughabghab and C. L. Dunford, BNL-NCS-65712, Conference 981003 (1998).
\bibitem{prakash} M. Prakash, J. J. Wambach, and Z. Y. Ma, Phys. Lett. B 128, 141 (1983).
\bibitem{jnde} J. N. De, S. Shlomo and S. K. Samaddar, Phys. Rev. C 57, 1398 (1998). 
\bibitem{hagel} K. Hagel et al., Nucl. Phys. A 486, 429 (1988).
\bibitem{gonin}  M. Gonin et al. , Phys. Lett. B 217, 406 (1989); Phys. Rev. C 42, 2125 (1990).
\bibitem{chibi} A. Chibihi et al., Phys. Rev. C 43, 652 (1991); Phys. Rev. C 43, 666 (1991).
\bibitem{fine} B. J. Fineman et al.,  Phys. Rev. C 50, 1991 (1994).
\bibitem{fabris} D. Fabris et al., Phys. Rev. C 50, R1261 (1994). 
\bibitem{coll_enhance} A. V. Ignatyuk, K. K. Istekov, and G. N. Smirenkin, Sov. J. Nucl. Phys. 29, 450 (1979).
\bibitem{proy1} Pratap Roy et al., Phys. Rev. C 88, 031601(R) (2013). 
\bibitem{kban1} K. Banerjee et al., Phys. Rev. C 85, 064310 (2012). 
\bibitem{proy2} Pratap Roy et al., Phys. Rev. C 86, 044622 (2012).
\bibitem{ykgupta1} Y. K. Gupta et al., Phys. Rev. C 80, 054611 (2009).
\bibitem{ykgupta2} Y. K. Gupta et al., Phys. Rev. Phys. Rev. C 78, 054609 (2008). 
\bibitem{m_agar} M. Aggarwal and S. Kailas, Phys. Rev. C 81, 047302 (2010).
\bibitem{mitra} A. Mitra, D.R. Chakrabarty, V.M. Datar, Suresh Kumar, E.T. Mirgule, H.H. Oza, V. Nanal, R.G. Pillay, 
Nucl. Phys. A 765, 277 (2006). 
\bibitem{kban} K. Banerjee et al., Nuclear Instruments and Methods in Physics Research Section A 608, 440 (2009).
\bibitem{deep} Deepak Pandit et al., Nuclear Instruments and Methods in Physics Research Section A 624, 148 (2010).
\bibitem{geant} S. Agostinelli, et al., Nuclear Instruments and Methods in Physics Research Section A 506, 250 (2003).
\bibitem{cas} F. Puhlhofer, et al., Nuclear Physics A 280, 267 (1976).
\bibitem{ign1} A. V. Ignatyuk, G. N. Smirenkin, and A. S. Tishin, Sov. J. Nucl. Phys. 21 (1975) 255 [Yad. Fiz. 21 (1975) 485].
\bibitem{per1} C.M. Perey and F.G. Perey, Atomic data nucl. data tables vol17, 1 (1976).
\bibitem{per2} F. G. Perey, Phys rev 131, 745 (1963).
\bibitem{sat}  L. Mcfadden and G. R. Satchler, Nucl. Phys. A 84, 177 (1966).
\bibitem{dang} N. Quang Hung  and N. Dinh Dang,  Phys. Rev. C 84, 054324 (2011).
\bibitem{man}  M. Gohil et al., EPJ Web Conferences 66, 03073 (2014) 
\end{thebibliography}
\end{document}